\documentstyle{article}

\begin{document}

\newtheorem{thm}{Theorem}[section]
\newtheorem{lemma}[thm]{Lemma}
\newtheorem{defin}[thm]{Definition}
\newtheorem{rem}[thm]{Remark}
\newtheorem{cor}[thm]{Corollary}

 %
 %
 % MMM MMM MMM MMM MMM
 %
 %

\def\fn{ \baselineskip = 0pt
\vbox{\hbox{\hspace*{3pt}\tiny $\circ$}\hbox{$f$}} \baselineskip = 12pt\!}
\def\gn{ \baselineskip = 0pt
\vbox{\hbox{\hspace*{2pt}\tiny $\circ$}\hbox{$g$}} \baselineskip = 12pt\!}
\newcommand{\lap}{\bigtriangleup}
\def\be{\begin{equation}}
\def\ee{\end{equation}}
\def\bea{\begin{eqnarray}}
\def\eea{\end{eqnarray}}
\def\beas{\begin{eqnarray*}}
\def\eeas{\end{eqnarray*}}

\def\dt{\partial_t}
\def\dx{\partial_x}
\def\dv{ \partial_v }
\def\R{{\rm I\kern-.1567em R}}
\def\N{{\rm I\kern-.1567em N}}
\def\Z{{\sf Z\kern-.3567em Z}}

\def\D{{\cal D}}
\def\s{\sigma}
\def\e{\epsilon}
\def\supp{\mbox{\rm supp}} 
\def\div{\mbox{\rm div}}
\def\n#1{\vert #1 \vert}
\def\nn#1{\Vert #1 \Vert}
\def\prf{\noindent
         {\bf Proof:\ }}
\def\prfe{\hspace*{\fill} $\Box$ 

\smallskip \noindent}

\title{Nonlinear stability of homogeneous models in Newtonian
       cosmology} 
\author{Gerhard Rein\\
        Mathematisches Institut \\
        der Universit\"at M\"unchen\\
        Theresienstr.\ 39\\
        80333 M\"unchen, Germany} 
\date{}
\maketitle

\begin{abstract}
We consider the Vlasov-Poisson system in a cosmological setting
as studied in \cite{RR2} and prove nonlinear stability
of homogeneous solutions against small, spatially periodic
perturbations in the $L^\infty$-norm of the spatial mass density. 
This result is connected with the question of how 
large scale structures such as galaxies have evolved
out of the homogeneous state of the early universe.  
\end{abstract}
 %
 %
 % INTRO
 %
 %
\section{Introduction}
In textbooks on astrophysics and cosmology
the formation of galaxies is sometimes explained by demonstrating
that for certain model equations spatially homogeneous solutions, 
which are
to represent the early state of the universe, are unstable
against small perturbations so that such perturbations grow and eventually
lead to the formation of large scale structures, cf.~\cite{Pe1,Pe2,We}. 

We intend to study the stability problem of such homogeneous
states in the context of the Vlasov-Poisson system:
\be \label{vl}
\dt f + v\cdot \dx f - \dx U \cdot \dv f =0,\ 
\ee
\be \label{po}
\lap U = 4\pi \rho,
\ee
\be \label{rho}
\rho (t,x) = \int f(t,x,v)\,dv .
\ee
Here $t\geq 0$ denotes time, $x\in \R^3$ position, and $v\in \R^3$ 
velocity,
$f=f(t,x,v)$ denotes the particle distribution function on phase space,
$\rho = \rho (t,x)$ denotes the spatial mass density generated by $f$,
and $U=U(t,x)$ denotes the gravitational potential.
Therefore, in this model the particles move according to Newton's
equations of motion
\[
\dot x = v,\ \dot v = -\dx U(t,x)
\]
under the influence
of the gravitational field which they create collectively
according to Newton's law of gravity, and collisions
as well as relativistic effects are neglected. 

The corresponding initial-value problem  where the phase space density
$f$ is prescribed  at $t=0$  is well understood for the case of an 
isolated system, i.~e., if the Vlasov-Poisson system
(\ref{vl}), (\ref{po}), (\ref{rho}) is supplemented
with the boundary condition $\lim_{x \to \infty} U(t,x) =0$
and a corresponding boundary condition for $f$, cf.\ 
\cite{Ho2,LP,Pf,Sch1}. Here we are interested in solutions
which model the evolution of a Newtonian universe, and for this situation 
these boundary conditions are not appropriate.
In the context of general relativity an isolated system can be thought
of as a localized deviation from flat space---an asymptotically flat
spacetime---whereas a cosmological solution is a deviation from
a homogeneous state, usually one with non-zero, spatially constant
mass density. Both situations have been investigated with the 
Vlasov equation as matter model under the assumption of special
symmetries, cf.~\cite{R3,RR1} and the references therein.
However, the existence theory for this so-called Vlasov-Einstein 
system is not yet sufficiently developed to form a basis
for the stability analysis intended here.

Spatially homogeneous, cosmological solutions for the Vlasov-Poisson 
system can be constructed
as follows: For a nonnegative, compactly supported function 
$H \in C^1_c (\R)$ we set
\be \label{fndef}
f_0 (t,x,v) := 
H\left( a^2(t) \Bigl|v - \frac{\dot a (t)}{a(t)} x\Bigr|^2 \right)
\ee
where $a$ is a positive, scalar function to be determined later. Then
\[
\rho_0(t,x) = \int f_0(t,x,v)\, dv = a^{-3}(t) \int H(v^2)\, dv ,
\]
and, after normalizing
\[
\int H(v^2)\, dv =1,
\]
we obtain the homogeneous mass density
\be \label{rndef}
\rho _0(t) = a^{-3}(t),\ t\geq 0.
\ee
A solution of the Poisson equation (\ref{po}) is then given by
\be \label{undef}
U_0(t,x) := \frac{2\pi}{3} a^{-3}(t)\, x^2,\ t\geq 0,\ x\in \R^3,
\ee
and it remains to determine the function $a$ in such a way
that $f_0$ satisfies the Vlasov equation (\ref{vl}) with force term
\[
\dx U_0(t,x) = \frac{4 \pi}{3} a^{-3}(t)\, x.
\]
A short computation shows that this is the case if
$a$ is a solution of the differential equation
\be \label{agl}
\ddot a + \frac{4 \pi}{3} a^{-2} =0 ,
\ee
which is the equation of radial motion in the gravitational field
of a point mass.
It is the purpose of the present paper to investigate the nonlinear
stability of such homogeneous models $(f_0,\rho_0,U_0)$ against small
perturbations. In order to have a meaningful concept of stability we
restrict ourselves to the case where this homogeneous solution
exists---and expands---for all time $t \geq 0$, which is the case if 
$\dot a (0) >0$ and
\[
E_a := \frac{1}{2} \dot a(t)^2 - \frac{4 \pi}{3} a^{-1} (t) >0,
\]
i.~e., the energy, a conserved quantity along solutions of (\ref{agl}), 
is positive.
If $E_a = 0$ the homogeneous solution also expands for all future time
but not sufficiently fast for our stability proof to go through.
We normalize the initial condition by requiring that $a(0)=1$;
the energy then is positive iff
\[
\dot a(0) > \sqrt{8\pi/3}.
\]
Note that the condition $a(0) =1$  means that the universe has already expanded out of
the initial big bang singularity $a=0$ for some time.
By the assumption on the energy there exist constants 
$0 < c_1 < c_2$ such that
\[
c_1 \leq \dot a (t) \leq c_2,\ t\geq 0.
\]
In order to study the stability of such a homogeneous model
we investigate the time evolution of small deviations from it,
that is, we consider solutions of the Vlasov-Poisson system
(\ref{vl}), (\ref{po}), (\ref{rho}) of the form
\[
f=f_0 + g,\ \rho =\rho_0 + \s,\ U = U_0 + W .
\]
To investigate the perturbations $g,\ \s,\ W$, which we want to assume spatially periodic, it is useful to perform the
following transformation of variables:
\be \label{tra}
\tilde x  =  a^{-1} (t) \,x,\quad 
\tilde v = v - a^{-1}(t) \dot a(t)\, x,\
\tilde g (t,\tilde x,\tilde v) = g(t,x,v) .
\ee
If we compute the system satisfied by $g,\ \s,\ W$ in these transformed
variables and afterwards drop the tildas we obtain the following
version of the Vlasov-Poisson system which governs the time evolution
of deviations from the homogeneous state:
\be \label{pvlt}
\dt  g + \frac{1}{a}  v\cdot \partial_{ x}  g - 
\frac{1}{a} \Bigl(\partial_{ x}  W + 
\dot a  v  \Bigr)\cdot \partial_{ v}  g =
2 a H'(a^2  v^2)\,  v \cdot \partial_{ x}  W, 
\ee
\be \label{ppot}
\lap  W = 4\pi a^2  \s,
\ee
\be \label{prhot}
 \s (t, x) = \int  g(t, x, v)\,dv .
\ee
It should be kept in mind that now $x,v \in \R^3$ are related
to the original variables by the transformation (\ref{tra}).
More details on this transformation
are given in \cite{RR2},
its necessity comes from the fact that in the
original variables the equations for the deviation $g$ would become
explicitly $x$-dependent which would exclude the possibility
of studying spatially periodic deviations, a class which seems physically
reasonable and is convenient for our mathematical analysis.

The main result of the present paper is that the homogeneous solutions
described above  
are nonlinearly stable against spatially periodic perturbations
in the following sense; for explanation of our notation and the set
of admissable data $\D$ we refer to the next section:
\begin{thm} \label{stab}
For every $\e_1 >0$ there exists $\e_0 >0$ such that every solution $g$
of $(\ref{pvlt}),\ (\ref{ppot}),\ (\ref{prhot})$ with $\gn \in \D$ and $\nn{\gn\,}_\infty < \e_0$
satisfies the estimate 
\[
\nn{\s (t)}_\infty < a^{-3} (t) \e_1,\ t \geq 0 .
\]
If we define the density contrast $\delta (t,x)$ of the corresponding
solution of $(\ref{vl})$,  $(\ref{po})$, $(\ref{rho})$ by
\[
\rho (t,x) = \rho_0 (t) \bigl( 1 + \delta (t,x) \bigr)
\]
then 
\[
\nn{\delta (t)}_\infty < \e_1 ,\ t \geq 0 .
\]
\end{thm}
Note that the result is really
one on Lyapunov stability and should not be
mistaken as an asymptotic stability result: the factor $a^{-3} (t)$
in the estimate for $\s$
simply reflects the overall expansion of the solution; the model 
contains no dispersive effects which would make the solution $f$ 
converge to $f_0$.

We briefly explain how the paper proceeds and how it relates to
other work in this field: An essential prerequisite for a stability
analysis is a global-in-time existence theorem  for
the corresponding initial value problem for the system 
(\ref{pvlt}), (\ref{ppot}), (\ref{prhot}). This was established
in \cite{RR2}. For easier reference we collect some
results and notation from that paper in the next section.
Stability and instability results for the Vlasov-Poisson system have 
been established for various, geometrically different settings
in the plasma physics case, cf.\ \cite{BR1,BMR,GS1,GS2,R2}. In the more
difficult stellar dynamics case linearized stability of
isolated systems is investigated in \cite{BMR}, and the question
of nonlinear stability was attacked in \cite{W} with dubious result,
cf.\ also \cite{Wo}. Our paper differs from these in that we
obtain a stability estimate with respect to the $L^\infty$-norm
of $\rho$ and obtain a rigorous, nonlinear result in the stellar
dynamics case. On the other hand, our techniques are more akin
to small data results known for the Vlasov-Poisson
and related systems such as the Vlasov-Maxwell and Vlasov-Einstein 
systems in the case of an isolated configuration, cf.\
\cite{BD,R1,RR1}. The essential new difficulty arises from
the fact that as opposed to these results we cannot use the
dispersive effect of the almost free streaming characteristics
for small fields in empty space,
since we are in a spatially periodic situation
with a non-zero background field. Instead, we make use of the expansion
of the background solution which is---for the case considered here---fast
enough to prevent any conglomeration and growth of small perturbations.
A related result, but for a different matter model, namely dust, is 
given in \cite{BrR}. However, note that there the authors have to 
assume a positive cosmological constant to have a sufficiently rapid expansion of the background solution to prevent growth of the 
perturbations. As a first step towards our result
we investigate in the third section how solutions, in particular their 
force field, depend on the initial data for $g$, that is, we prove continuous dependence on the initial data in an appropriate sense.
Next we investigate how the characteristics
behave under a small perturbation of the background field and
establish certain resulting fall-off estimates for the spatial
mass density of the perturbation.  In the fifth section
we combine these two sets of estimates in a bootstrap argument
to prove Theorem~\ref{stab}: By making the initial data small and using 
the continuous dependence of the field on the initial data we make sure
that the field is a small perturbation of the background field in the 
sense of
Section 4 so that by the results of that section we get new fall-off
estimates on the perturbed field which turn out to be asymptotically stronger than the smallness assumption we start with. 
The time interval on which these fall-off estimates hold can therefore 
be extended to all of $[0,\infty[$, and the proof is complete.
Finally, we will briefly
discuss how our result relates to the usual textbook 
explanation of the formation of galaxies. If this explanation
is mathematically false, then why do galaxies exist?

 %
 % NOTATION
 % 
\section{Notation and preliminary results}
\setcounter{equation}{0}

Let $Q:= [0,1]^3$ and $S:= Q \times \R^3$. We need the following
spaces of periodic functions:
\beas
{\cal P} (Q)&:=& \Bigl\{ h: \R^3 \to \R \mid h(x + \alpha ) = h(x),\
x\in \R^3,\ \alpha \in \Z^3 \Bigr\} ,\\
{\cal P} (S)&:=& \Bigl\{ h: \R^6 \to \R \mid h(x + \alpha ,v ) = 
h(x,v),\ x, v\in \R^3,\ \alpha \in \Z^3 \Bigr\}  , \\
C^n_\pi (Q) &:=& C^n (\R^3) \cap {\cal P }(Q), \\
C^n_\pi (S) &:=& C^n (\R^6) \cap {\cal P }(S), \\
C^n_{\pi,c} (S) &:=& \Bigl\{h \in C^n_\pi (S) \mid
\exists u\geq 0 : h(x,v)=0,\ \n{v} >u \Bigr\} .
\eeas
For $p\in [1,\infty]$ we denote by $\nn{\cdot}_p $ the usual $L^p$-norm,
where the integral (or supremum) extends over $Q$ or $S$ as the case may 
be, and
\[
L^p_\pi (Q) := L^p (Q) \cap {\cal P} (Q) ,\
L^p_\pi (S) := L^p (S) \cap {\cal P} (S) ,
\]
where we identify functions in ${\cal P} (Q)$ and ${\cal P} (S)$
with their restrictions to $Q$ or $S$ respectively.
For easier reference we collect some results from \cite{RR2}:
\begin{thm} \label{glex}
For every initial datum $\gn \in C^1_{c,\pi} (S)$ with $\int_S \gn =0$ 
and $\gn (x,v) + H(v^2) \geq 0$ for $x, v\in \R^3$ there exists
a unique solution $g \in C^1 ([0,\infty[ \times \R^6) \cap C([0,\infty[;
C^1_\pi (S))$ of the system {\rm (\ref{pvlt}), (\ref{ppot}), 
(\ref{prhot})} with $g(0)= \gn$. 
\end{thm} 
This is Thm.~6.1 of \cite{RR2}; note that in the present case
the homogeneous background solution exists for all $t \geq 0$.
The restriction $\int_S \gn =0$ means that the perturbation does not
change the total mass of the background solution and is necessary
in order to allow for spatially periodic solutions of the Poisson
equation (\ref{ppot}),
$\gn (x,v) + H(v^2) \geq 0$ means that $f(0)$ is a nonnegative function.

Next we have the following representation of $g$ in terms of the
characteristics of (\ref{pvlt}), cf.\ \cite[Lemma 4.3 and Lemma 4.4]{RR2}:
\begin{lemma} \label{chasol}
Let $(X,V)(s,t,x,v)$ denote the solution of the characteristic
system
\be \label{cha}
\dot x = \frac{1}{a(s)} v,\ 
\dot v = - \frac{1}{a(s)} \Bigl( \dx W (s,x) + \dot a(s) v \Bigr) ,
\ee
with $(X,V)(t,t,x,v) = (x,v)$, $x, v \in \R^3$, $0 \leq s \leq t$.
Then 
\[
g(t,x,v) =
\gn \bigl((X,V)(0,t,x,v)\bigr) + H\bigl(V^2 (0,t,x,v)\bigr) - 
H\bigl(a^2(t) v^2\bigr),\ t\geq 0,\ x, v \in \R^3 .
\]
\end{lemma}
Finally, we need the following estimates on the solutions of the Poisson
equation in the spatially periodic set-up:
\begin{lemma} \label{green}
For $\sigma \in C^1_\pi (Q)$ with $\int_Q \s (x)\, dx =0$ there exists
in $C^2_\pi (Q)$ a unique solution of
\[
\lap W = 4 \pi \s
\]
with $\int_Q W(x)\, dx =0$. It satisfies the following estimates:
\bea \label{fiest}
\nn{\dx W}_\infty 
&\leq& 
C \left( \nn{\s}_1^{1/3} \nn{\s}_\infty^{2/3}
+ \nn{\s}_1 \right) ,\\
\nn{\dx ^2 W}_\infty 
&\leq&
C \Bigl(\bigl(1 + \nn{\s}_\infty\bigr)
\bigl(1 + \ln ^\ast \nn{\dx \s}_\infty/2 \bigr) +\nn{\s}_1 \Bigr),   \label{dfiest1}\\
\nn{\dx ^2 W}_\infty 
&\leq&
C \left(\nn{\s}_1^{1/16} \nn{\s}_\infty^{3/4} \nn{\dx \s}_\infty^{3/16}
+ \nn{\s}_1 \right), \label{dfiest2}
\eea 
where
\[
\ln^\ast s := \left\{ \begin{array}{ccc}
s &,& s \leq 1\\
1 + \ln s &,& s > 1
\end{array} \right. .
\]
\end{lemma}
Except for the estimate (\ref{dfiest2}) this is contained in
\cite[Lemma 4.1 and Lemma 4.2]{RR2}, the estimate (\ref{dfiest2})
follows from \cite[Lemma 1]{BD} and the fact that the singularity
in the Green's function for the present, spatially periodic
situation is the same as for the problem on the whole space,
cf.~\cite[Lemma 4.1]{RR2}. 

In what follows we consider only small perturbations $\gn$ which
will be taken from the set 
\beas
\D := \biggl\{ \gn \in C^1_{\pi,c} (S)
&\mid&
\int_S \gn =0,\ \gn (x,v) + H(v^2) \geq 0 \ \mbox{for all}\ 
x,v \in \R^3,\\
&&
\nn{\gn\,}_{1,\infty} < d_0,\ \mbox{and}\ \gn (x,v) = 0\ \mbox{for}\
\n{v} \geq u_0 \biggr\}.
\eeas
Here $d_0 >0$ and $u_0 >0$ are some fixed constants,
and  $\nn{\gn\,}_{1,\infty}$ denotes
the $L^\infty$-norm of $\gn$ and its first order derivatives. 
Note that in Theorem~\ref{stab} we need to make $\nn{\gn\,}_\infty$
small but we also need a uniform bound on $\nn{\gn\,}_{1,\infty}$.
In order to keep
our notation simple we use the following \newline
{\bf Convention on constants:} Constants denoted by $C$ may only depend
on the initial data set $\D$, on $c_1$ and $c_2$ determined by
$0 < c_1 \leq \dot a (t) \leq c_2 <\infty,\ t \geq 0$, and, starting
from Section~4, also on the parameter $\delta$ introduced there,
and they may change from line to line. The same convention applies
to functions $C \in C([0,\infty[)$.

 %
 % CONDEP
 %
\section{Dependence of the field on the initial data}
\setcounter{equation}{0}

The purpose of this section is to show that we can make the perturbed
field and its spatial derivative small on any given, finite time 
interval by making
$\nn{\gn\,}_\infty$ small. As a first step in this
direction we establish this result for the perturbed density and field:
\begin{lemma} \label{condep1}
There exists a nondecreasing function $C \in C([0,\infty[)$ 
such that for every solution of $(\ref{pvlt}),\
(\ref{ppot}),\ (\ref{prhot})$ with initial condition $\gn \in \D$
the estimate
\[
\nn{\s (t)}_1 + \nn{\s (t)}_\infty + \nn{\dx W(t)}_\infty  \leq C(t)\, \nn{\gn\,}_\infty
\]
holds for $t \geq 0$.
\end{lemma}
\prf
There exists a nondecreasing function $P \in C([0,\infty[)$ such that
\be \label{vest} 
g(t,x,v) = 0,\ \n{v} \geq P(t),\ x \in \R^3,\ t \geq 0
\ee
for every solution with initial condition $\gn \in \D$.
This follows from \cite[Thm.~6.1]{RR2}; that this function can be 
chosen uniformly on the set $\D$ follows from the fact that all
parameters which enter into its construction are uniformly bounded
on $\D$. Lemma~\ref{chasol} and the mean value theorem imply that
\[
\nn{g(t)}_\infty \leq \nn{\gn\,}_\infty + C \sup_{x,v \in \R^3}
\bigl| V(0,t,x,v) - a(t) v\bigr| .
\]
Since
\[
\frac{d}{ds} \Bigl( a(s) V(s,t,x,v) \Bigr)
= - \dx W\bigl(s,X(s,t,x,v)\bigr)
\]
it follows that
\[
\nn{g(t)}_\infty \leq \nn{\gn\,}_\infty + 
C \int_0^t \nn{\dx W(s)}_\infty ds.
\]
With (\ref{vest}) we conclude that
\be \label{sestinf}
\nn{\s (t)}_\infty \leq C(t) \left(\nn{\gn\,}_\infty + 
\int_0^t \nn{\dx W(s)}_\infty ds \right)
\ee
and
\be \label{sest1}
\nn{\s (t)}_1 \leq C(t) \left(\nn{\gn\,}_\infty + 
\int_0^t \nn{\dx W(s)}_\infty ds \right) .
\ee
Inserting these estimates into  (\ref{fiest})
yields the Gronwall type inequality
\[
\nn{\dx W(t)}_\infty \leq C(t) \left(\nn{\gn\,}_\infty + 
\int_0^t \nn{\dx W(s)}_\infty ds \right)
\]
from which the estimate
\[
\nn{\dx W(t)}_\infty \leq C(t) \nn{\gn\,}_\infty,\ t \geq 0,
\]
follows. If we insert this into
(\ref{sestinf}) and (\ref{sest1}) the proof is complete. \prfe
Next we want to prove an analogous estimate for the $x$-derivatives
of the perturbed field. To do so we need certain estimates on the
$x$-derivatives of the characteristics which will follow from the
following slightly rewritten form of the characteristic
system (\ref{cha}): If
\[
y(s) := a(s) X(s,t,x,v)
\]
for $t,x,v$ fixed then $y$ satisfies the differential
equation
\be \label{chay}
\ddot y = - \frac{4\pi}{3} \frac{1}{a^3 (s)}\, y 
- \frac{1}{a (s)} \,\dx W \bigl(s, y/a(s)\bigr)
\ee
where we have used (\ref{agl}) to express $\ddot a$.
\begin{lemma} \label{xdercha}
There exists a constant $C$ such that every solution of $(\ref{cha})$
satisfies the estimates
\beas
\left| \dx X(s,t,x,v) \right|
&\leq&
C \frac{a(t)}{a(s)} \exp \left( C \int_s^t a^{-1} (\tau)\,
\nn{\dx^2 W(\tau)}_\infty d\tau \right),\\
\left| \dx V(s,t,x,v) \right|
&\leq&
\frac{1}{a(s)}   \int_s^t  \nn{\dx^2 W(\tau)}_\infty 
\left| \dx X(\tau ,t,x,v) \right|\,d\tau
\eeas
for $x,v \in \R^3$ and $0\leq s \leq t$.
\end{lemma}
\prf
Let $\xi (s) = a(s) \dx X(s,t,x,v) = \dx y(s)$. Then by (\ref{chay}) 
we have
\[
\ddot \xi =  - \frac{4\pi}{3} \frac{1}{a^3 (s)}\, \xi 
- \frac{1}{a^2 (s)}\, 
\dx^2 W \bigl(s, X(s,t,x,v)\bigr) \, \xi,
\]
and
\[
\xi (t) = a(t)\, {\rm id},\ \dot \xi (t) = \dot a(t)\, {\rm id} .
\]
Thus
\[
\n{\dot \xi (s)}
\leq
c_2 + \int_s^t \left( \frac{4\pi}{3} a^{-3}(\tau) + a^{-2} (\tau) \,  
\nn{\dx^2 W(\tau)}_\infty\right)  \n{\xi (\tau)}\, d\tau ,
\]
and
\beas
\n{\xi (s)}
&\leq&
a(t) + c_2 (t-s) +
\int_s^t \int_\s^t \left( C a^{-3}(\tau) + a^{-2} (\tau) \, 
\nn{\dx^2 W(\tau)}_\infty \right) \n{\xi (\tau)}\, d\tau\, d\s\\
&\leq&
C a(t) +
\int_s^t \int_s^\tau d\s \left( C a^{-3}(\tau) + a^{-2} (\tau)\, 
\nn{\dx^2 W(\tau)}_\infty \right) \n{\xi (\tau)}\, d\tau \\
&\leq&
C a(t) +
\int_s^t \tau \left( C a^{-3} (\tau) + a^{-2} (\tau)   
\nn{\dx^2 W(\tau)}_\infty \right)
\n{\xi (\tau)}\, d\tau .
\eeas
Now a Gronwall argument yields
\beas
\n{\xi (s)} 
&\leq&
C a(t) \exp \left( C \int_s^t \tau a^{-3} (\tau)\, d\tau \right)
\exp \left( \int_s^t \tau a^{-2} (\tau)\,
\nn{\dx^2 W(\tau)}_\infty d\tau \right)\\
&\leq&
C a(t) \exp \left( C \int_s^t a^{-1} (\tau)\,
\nn{\dx^2 W(\tau)}_\infty d\tau \right),
\eeas
which is the first estimate of the lemma. Next observe that
\be \label{veq}
V(s,t,x,v) = \frac{a(t)}{a(s)} v + \frac{1}{a(s)}
\int_s^t \dx W \bigl(\tau, X(\tau,t,x,v)\bigr)\, d\tau;
\ee
this is the solution of the initial value problem
\[
\dot V = - \frac{\dot a (s)}{a(s)} V - \frac{1}{a(s)} 
\dx W\bigl(s,X(s,t,x,v)\bigr),
\ V(t) =v .
\]
Therefore,
\[
\dx V(s,t,x,v) =  \frac{1}{a(s)}
\int_s^t \dx^2 W \bigl(\tau, X(\tau,t,x,v)\bigr)\, \dx 
X(\tau,t,x,v)\, d\tau
\]
from which the second estimate of the lemma follows. \prfe
\begin{lemma} \label{condep2}
There exists a nondecreasing function $C \in C([0,\infty[)$ 
such that for every solution of $(\ref{pvlt}),\
(\ref{ppot}),\ (\ref{prhot})$ with initial condition $\gn \in \D$
the estimate
\[
\nn{\dx^2 W(t)}_\infty  \leq C (t)\, \nn{\gn\,}_\infty^{13/16}
\]
holds for $t \geq 0$.
\end{lemma}
\prf
Lemma~\ref{chasol} implies that for $\gn \in \D$,
\beas
\nn{\dx g(t)}_\infty 
&\leq&
d_0
\sup_{x,v \in \R^3} \Bigl( \n{\dx X(0,t,x,v)} + \n{\dx V(0,t,x,v)} 
\Bigr)\\
&&
+ \, C \sup_{x,v \in \R^3} \n{\dx V(0,t,x,v)}.
\eeas
From Lemma~\ref{xdercha} it follows that
\[
\bigl| \dx X(0,t,x,v) \bigr|
\leq C a(t) \exp\left( C \int_0^t \nn{\dx^2 W(\tau)}_\infty d\tau \right)
\]
and 
\beas
\bigl| \dx V(0,t,x,v) \bigr|
&\leq&
C \int_0^t \nn{\dx^2 W(s)}_\infty \frac{a(t)}{a(s)} 
\exp\left( C \int_s^t a^{-1} (\tau)\,\nn{\dx^2  W(\tau)}_\infty d\tau \right)\,ds \\
&\leq&
C a(t) \int_0^t \nn{\dx^2 W(s)}_\infty  
\exp\left( C \int_s^t \nn{\dx^2 W(\tau)}_\infty d\tau \right) \,ds \\
&\leq&
C a(t) \exp \left( C \int_0^t \nn{\dx^2 W(\tau)}_\infty d\tau \right) .
\eeas
Therefore,
\[
\nn{\dx g(t)}_\infty  
\leq
C (t) \exp\left( C \int_0^t\nn{\dx^2 W(\tau)}_\infty d\tau \right),
\]
and by (\ref{vest}),
\[
\nn{\dx \s (t)}_\infty  
\leq
C (t) \exp\left( C \int_0^t\nn{\dx^2 W(\tau)}_\infty d\tau \right).
\] 
If we insert this estimate into (\ref{dfiest1}) and observe that
$\nn{\s (t)}_1$ and $\nn{\s (t)}_\infty$ can be bounded by some
$C(t)$, cf.~Lemma~\ref{condep1}, 
we obtain the Gronwall type inequality
\beas
\nn{\dx^2 W (t)}_\infty  
&\leq&
C(t) + C(t)
\ln^\ast \Biggl[ C (t) \exp\left( C \int_0^t\nn{\dx^2 
W(\tau)}_\infty d\tau \right) \Biggr] \\ 
&\leq&  
C(t)  + C(t)\, \int_0^t \nn{\dx^2 W(\tau)}_\infty d\tau .
\eeas
A Gronwall argument implies that there exists a function
$C \in C([0,\infty[)$ such that for solutions with $\gn \in \D$ the
estimate
\[
\nn{\dx^2 W (t)}_\infty \leq C(t) ,\ t \geq 0,
\]
and thus also 
\[
\nn{\dx \s (t)}_\infty \leq C(t) ,\ t \geq 0,
\]
holds. If we now insert this estimate and the estimates from 
Lemma~\ref{condep1} into (\ref{dfiest2}) the assertion of the 
lemma follows.
\prfe
 
 %
 % FREE STREAMING
 %
\section{The free streaming condition and its consequences}
\setcounter{equation}{0}

Although this terminology is not quite correct we refer in the following
to the characteristics of the homogeneous background solution
as the free streaming characteristics.
We need to make precise what it means that a solution behaves almost 
like the homogeneous background solution. To this end we fix a
constant $\delta \in ]0,13/16[$. For $T>0$ and a parameter $\gamma \in ]0,1]$
we say that a solution $g$ of (\ref{pvlt}), (\ref{ppot}),\ (\ref{prhot})
satisfies the free streaming condition on the time interval $[0,T]$
with parameter $\gamma >0$ if
\[
{\rm (FS\gamma)} \qquad \qquad
\renewcommand{\arraystretch}{2}
\left\{
\begin{array}{c}
\displaystyle
\nn{\dx W (t)}_\infty \leq \gamma a^{-\delta} (t)\\
\displaystyle
\nn{\dx^2 W (t)}_\infty \leq \gamma a^{-\delta} (t)
\end{array} \right. ,\ 0 \leq t \leq T .
\renewcommand{\arraystretch}{1} 
\]   
We want to study the behaviour of the characteristics
of (\ref{pvlt}) under the assumption (FS$\gamma$).
As a first step we estimate the derivatives considered in
Lemma~\ref{xdercha} under the assumption (FS$\gamma$):
\begin{lemma} \label{xderchafs}
Let the condition {\rm (FS$\gamma$)} hold for some $\gamma \in ]0,1]$
on some time interval $[0,T]$. Then
\[
\left| \dx X(s,t,x,v) \right| \leq
C \frac{a(t)}{a(s)} ,\
\left| \dx V(s,t,x,v) \right| \leq 
C \gamma \frac{a(t)}{a(s)} ,\  0 \leq s \leq t\leq T ,\ x,v \in \R^3.
\]
\end{lemma}
\prf
By (FS$\gamma$) and since $a(\tau) \geq 1 + c_1 \tau$ and $\gamma \leq 1$, 
\[
\int_s^t a^{-1} (\tau)\, \nn{\dx^2 W(\tau)}_\infty d\tau 
<
\int_0^\infty a^{-1-\delta} (\tau)\, d\tau < \infty .
\]
Inserting this
estimate into the first estimate in Lemma~\ref{xdercha}
proves the first assertion, and inserting the latter into the second
estimate in Lemma~\ref{xdercha} yields
\[
\left| \dx V(s,t,x,v) \right|
\leq \frac{1}{a(s)} C \int_s^t \gamma a^{-\delta}(\tau) 
\frac{a(t)}{a(\tau)}
d\tau < C \frac{a(t)}{a(s)} \gamma \int_0^\infty a^{-1-\delta} (\tau)\,d\tau,
\]
and the proof is complete. \prfe

\smallskip
\noindent
{\bf Remark:} Without the normalizing condition $a(0) =1$ the 
constant $C$ in Lemma~\ref{xderchafs} would depend on $a(0)$ in 
such a way that $C \to \infty$
as $a(0) \to 0$. We will come back to this observation at the end of the last section.   

\begin{lemma} \label{dvvfs}
Let the condition {\rm (FS$\gamma$)} hold for $\gamma \in ]0,1]$
on some time interval $[0,T]$. Then the estimate
\[
\left| \frac{1}{a(t)} \dv V(0,t,x,v) - {\rm id} \right|
\leq C \, \gamma 
\]
holds for $x,v \in \R^3$ and $t \in [0,T]$. If $\gamma \in ]0,1]$ 
is small enough then the mapping 
\[
\R^3 \ni v \mapsto V(0,t,x,v) \in \R^3
\]
is one-to-one, and
\[
\bigl| \det \dv V(0,t,x,v) \bigr| > \frac{1}{2} a^3 (t) .
\]
\end{lemma}
\prf
Let
\[
\xi (s) := a(s)\, \dv X(s,t,x,v) .
\]
Then
\[
\ddot \xi = - \frac{4 \pi}{3} \frac{1}{ a^3 (s)} \xi - \frac{1}{ a^2(s)}
\dx^2 W(s,x(s))\, \xi ,
\]
where $x(s) = X(s,t,x,v)$, cf.~(\ref{chay}).
In the free streaming case $\dx^2 W =0$ the solution takes the form 
\[
\xi_F (s) = - a(s) \, a(t) \int_s^t \frac{d\tau}{a^2 (\tau)} {\rm id};
\]
this is the solution of the initial value problem
\[
\ddot \xi_F = - \frac{4 \pi}{3} \frac{1}{ a^3 (s)} \xi_F,\
\xi_F (t) = 0,\ \dot \xi_F (t) = {\rm id}.
\]
Define
\[
\zeta (s) := \xi (s) - \xi_F (s) .
\]
Then
\[
\zeta (t) = \dot \zeta (t) = 0,
\]
and
\[
\ddot \zeta = - \frac{4 \pi}{3} \frac{1}{ a^3 (s)} \zeta 
-  \frac{1}{ a^2(s)} \dx^2 W\bigl(s,x(s)\bigr) \zeta 
-  \frac{1}{ a^2(s)} \dx^2 W\bigl(s,x(s)\bigr) \xi_F.
\]
Therefore,
\beas
\n{\ddot \zeta (s)} 
&\leq&
C a^{-2 -\delta} (s) \n{\zeta (s)} + C a(t) a^{-1-\delta} (s)
\int_s^t \frac{d\tau}{a^2(\tau)}\\
&\leq&
C a^{-2 -\delta} (s) \n{\zeta (s)} + C a(t) a^{-2 - \delta/2} (s)
\int_s^t a^{-1-\delta/2}(\tau)\, d\tau\\
&\leq&
C a^{-2 -\delta} (s) \n{\zeta (s)} + C a(t) a^{-2 - \delta/2} (s)
\eeas
and
\[
\n{\dot \zeta (s)} \leq
C a(t) \int_s^t a^{-2 -\delta/2} (\s) d\s
+ C \int_s^t a^{-2 -\delta}(\tau)\, \n{\zeta (\tau)}\, d\tau ,
\]
\beas
\n{\zeta(s)}
&\leq&
C a(t)  \int_s^t \int_\s^t a^{-2 -\delta/2} (\tau) d\tau\, d\s
+  C  \int_s^t \int_\s^t
a^{-2 -\delta}(\tau) \n{\zeta (\tau)}\, d\tau\, d\s\\
&=&
C a(t)  \int_s^t \int_s^\tau d\s a^{-2 -\delta/2} (\tau) d\tau
+  C \int_s^t \int_s^\tau d\s
a^{-2 -\delta}(\tau) \n{\zeta (\tau)}\, d\tau\\
&\leq&
C a(t)  \int_s^t \tau a^{-2 -\delta/2} (\tau) d\tau 
+  C \int_s^t \tau a^{-2 -\delta}(\tau) \n{\zeta (\tau)} \, d\tau \\
&\leq&
C\, a(t) + C\, \int_s^t a^{-1-\delta}(\tau)\,
\n{\zeta(\tau)}\, d\tau,
\eeas
from which we conclude that
\[
\n{\zeta (s)} \leq C\, a(t) 
\exp \left( C \int_s^t a^{-1-\delta}(\tau)\,\, d\tau\right)
\leq C\, a(t),
\]
i.~e.,
\[
\bigl| \dv X(s,t,x,v) \bigr| \leq C \frac{a(t)}{a(s)}  
+ a(t) \int_s^t \frac{d\tau}{a^2 (\tau)} .
\]
From (\ref{veq}) we obtain
\[
\dv V(s,t,x,v) = \frac{a(t)}{a(s)} {\rm id} + \frac{1}{a(s)}
\int_s^t \dx^2 W\bigl(\tau, x(\tau)\bigr)\, \dv X(\tau, t,x,v)\, d\tau,
\]
and thus
\beas
\bigl| \dv V(0,t,x,v) - a(t)\,{\rm id} \bigr|
&\leq&
\gamma \int_0^t a^{-\delta} (\tau) \left( C \frac{a(t)}{a(\tau)} 
+ a(t) \int_\tau^t \frac{d\s}{a^2(\s)}\right)d\tau\\
&\leq&
C \gamma a(t) \left(
\int_0^t a^{-1 -\delta} (\tau)\, d\tau + \int_0^t a^{-\delta}(\tau)
\int_\tau^t \frac{d\s}{a^2(\s)} d\tau\right)\\
&\leq&
C \gamma a(t) \left( 1 + \int_0^t a^{-1 - \delta/2}
(\tau)\int_\tau^t a^{-1-\delta/2} (\s) d\s d\tau \right)\\
&\leq&
C \gamma a(t) ,
\eeas
i.~e.,
\[
\left| \frac{1}{a(t)} \dv V(0,t,x,v) - {\rm id} \right| \leq C \gamma,
\]
an estimate which holds for all $t \in [0,T]$, $x,v \in \R^3$,
and $\gamma \in ]0,1]$. By making $\gamma$ small and using the mean
value theorem we conclude that
the mapping $v \mapsto V(0,t,x,v)$ is one-to-one and 
\[
\left| \det \frac{1}{a(t)} \dv V(0,t,x,v) \right| > \frac{1}{2}
\]
for all $x,v \in \R^3$ and $t \in [0,T]$. \prfe
Using the estimates of the previous lemma we can now show 
that the spatial mass density of the perturbation decays in time:
\begin{lemma} \label{sigdec}
Let $\gamma$ be as in Lemma~\ref{dvvfs} and let $g$ be a solution
with $\gn \in \D$, which satisfies {\rm (FS$\gamma$)} on some
time interval $[0,T]$. Then the estimates
\[
\nn{\s (t)}_\infty + \nn{\s (t)}_1 \leq
C\, a^{-3} (t) \bigl( \nn{\gn\,}_\infty + \gamma \bigr) 
\]
and
\[
\nn{\dx \s (t)}_\infty \leq
C\, a^{-2} (t) \bigl( \nn{\gn\,}_{1,\infty} + \gamma \bigr) 
\]
hold on $[0,T]$.
\end{lemma}
\prf
We use the transformation of variables
\[
v \mapsto V = V(0,t,x,v)
\]
together with the representation for $g$ from Lemma~\ref{chasol}
and the estimate in Lemma~\ref{dvvfs} to obtain
\beas
\n{\s (t,x)} 
&\leq&
\int \Bigl| \gn \bigl(X(0,t,x,v),V(0,t,x,v)\bigr)\Bigr|\, dv\\
&&
+ \left| \int H\bigl(V^2 (0,t,x,v)\bigr)\, dv - 
\int H\bigl(a^2(t)v^2\bigr)\, dv \right|\\
&\leq&
\int \Bigl| \gn \bigl(X(0,t,x,v),V\bigr)\Bigr| 
\bigl|\det \dv V(0,t,x,v) \bigr|^{-1} dV \\
&&
+  \left| \int H(V^2)\, \bigl|\det \dv V(0,t,x,v) \bigr|^{-1}dV 
- a^{-3}(t) \int H(V^2)\, dV \right|\\ 
&\leq&
C\, a^{-3}(t) \nn{\gn\,}_\infty + \int H(V^2)
\left| \frac{1}{\n{\det \dv V(0,t,x,v)}} - \frac{1}{a^3 (t)} \right| dV .
\eeas
Again by Lemma~\ref{dvvfs} we obtain the estimate
\beas
\left| \frac{1}{\n{\det \dv V(0,t,x,v)}} - \frac{1}{a^3 (t)} \right|
&\leq&
\frac{1}{a^3 (t) \n{\det \dv V(0,t,x,v)}}\\
&&
\cdot \bigl| \det \dv V(0,t,x,v) - \det (a (t) {\rm id})\bigr|\\
&\leq&
C a^{-6} (t) C a^2 (t) \bigl| \dv V(0,t,x,v) - a (t) {\rm id}\bigr|\\
&\leq&
C a^{-3} (t)\, \gamma ,
\eeas
and inserting this into the above estimate for $\s$ we have
\[
\nn{\s (t)}_\infty \leq C\, a^{-3}(t) \bigl( \nn{\gn\,}_\infty + \gamma \bigr).
\]
Since the $x$-integration extends only over the cube $Q$, the
same estimate holds for $\nn{\s (t)}_1$.
To obtain the estimate for $\dx \s$ we differentiate
the formula for $g$ in Lemma~\ref{chasol}, integrate with respect to
$v$, and use the same transformation of variables as above
together with the estimates from Lemma~\ref{xderchafs} and \ref{dvvfs}:
\beas
\n{\dx \s (t,x)}
&\leq&
C a(t) \int \Bigl| \bigl(\partial_{(x,v)} \gn \bigr) \bigl((X,V)(0,t,x,v)\bigr)\Bigr|\, dv\\
&&
+ C \gamma a(t) \int \left| \dv H (v^2)_{|v = V(0,t,x,v)}\right|\, dv\\
&\leq&
C a^{-2}(t) \bigl( \nn{\gn\,}_{1,\infty} + \gamma \bigr),
\eeas
and the proof is complete. \prfe 
%
%

 %
 % BOOTSTRAPP
 %
\section{The bootstrap argument}
\setcounter{equation}{0}
The purpose of this section is to prove Theorem~\ref{stab}.
To do so we first note that under the assumption (FS$\gamma$)
we obtain an asymptotically stronger decay-in-time for the perturbed 
field than the one defining (FS$\gamma$):
\begin{lemma} \label{fidec}
Let $\gamma$ be as in Lemma~\ref{dvvfs} and let $g$ be a solution
with $\gn \in \D$, which satisfies {\rm (FS$\gamma$)} on some
time interval $[0,T]$. Then the estimates
\[
\nn{\dx W (t)}_\infty \leq
C\, a^{-1} (t)  
\]
and
\[
\nn{\dx^2 W (t)}_\infty \leq
C\, a^{-13/16} (t)  
\]
hold on $[0,T]$.
\end{lemma}
\prf
By Lemma~\ref{green} we have
\[
\nn{\dx W (t)}_\infty \leq 
C a^2 (t) \left( \nn{\s (t)}_1^{1/3} \nn{\s (t)}_\infty^{2/3} + \nn{\s (t)}_1 \right)
\]
and
\[
\nn{\dx ^2 W(t)}_\infty \leq
C a^2 (t) \left(\nn{\s (t)}_1^{1/16} \nn{\s (t)}_\infty^{3/4} 
\nn{\dx \s (t)}_\infty^{3/16}
+ \nn{\s (t)}_1 \right); 
\]
note that the right hand side of the Poisson equation (\ref{ppot})
contains the factor $a^2(t)$. If we insert the estimates from
Lemma~\ref{sigdec} the assertion follows. \prfe
{\bf Proof of Theorem~\ref{stab}:}
Let $\epsilon_1 >0$ and choose 
\[
\gamma < \frac{\epsilon_1}{2C_1}
\]
in such a way, that Lemma~\ref{dvvfs} holds; $C_1$ denotes the constant
from Lemma~\ref{sigdec}. Next choose $T>0$ such that
\[
C_2 a^{-1} (t) \leq \frac{\gamma}{2} a^{-\delta} (t),\
C_2 a^{-13/16} (t) \leq \frac{\gamma}{2} a^{-\delta} (t),\ t \geq T,
\]
where $C_2$ denotes the constant from Lemma~\ref{fidec}. Finally, choose
\[
\epsilon_0 < \frac{\epsilon_1}{2 C_1}
\]
in such a way that
\[
C_3 (T) \epsilon_0 + C_4 (T) \epsilon_0^{13/16} < \frac{\gamma}{2} a^{-\delta} (T)
\]
where $C_3,\ C_4$ are the functions constructed in Lemma~\ref{condep1} 
and Lemma~\ref{condep2} respectively. 
Let $g$ be a solution with initial condition
$\gn \in \D$ such that $\nn{\gn\,}_\infty <\epsilon_0$.
The functions $C_3$ and $C_4$
are nondecreasing and $a$ is increasing. Therefore, by construction and Lemma~\ref{condep1} and \ref{condep2}
the free-streaming condition (FS$\gamma$) with parameter $\gamma$ holds
on some interval $[0,T^\ast[$ with $T^\ast >T$; we choose $T^\ast \in ]T,\infty]$ maximal with this property. By Lemma~\ref{fidec} the
estimates
\[
\nn{\dx W (t)}_\infty \leq
C_2\, a^{-1} (t)  
\]
and
\[
\nn{\dx^2 W (t)}_\infty \leq
C_2\, a^{-13/16} (t)  
\]
hold on $[0,T^\ast[$, and the choice of $T$ implies that
the free-streaming condition holds on $[T,T^\ast[$ with parameter
$\gamma/2$. Since $T^\ast$ is chosen maximal such that 
(FS$\gamma$) holds on $[0,T^\ast[$ it follows that
$T^\ast = \infty$. In particular,
\[
\nn{\s (t)}_\infty \leq
C_1\, a^{-3} (t) \bigl( \nn{\gn\,}_\infty + \gamma \bigr) 
< \epsilon_1 a^{-3} (t)
\]
by Lemma~\ref{sigdec} and by the choice of $\gamma$ and $\epsilon_0$.
It remains to see that this estimate implies the estimate on the 
density contrast $\delta (t,x)$. Let $(f,\rho,U)$ be the solution
of the Vlasov-Poisson system (\ref{vl}), (\ref{po}), (\ref{rho})
corresponding to $(g,\s,W)$, in particular,
\[
\rho (t,x) = \rho_0 (t) + \s (t,x) = \rho_0 (t) \bigl( 1 + a^3 (t) 
\s (t,x) \bigr) .
\]
Thus
\[
\delta (t,x) = a^3 (t) \s (t,x)
\]
where $(g, \s, W)$ is the perturbation written in the original
variables, i.~e., before applying the transformation (\ref{tra}).
If we denote the perturbation in the transformed variables
by $(\tilde g,\tilde \s, \tilde W)$ then by (\ref{tra}),
\[
\tilde \s (t,\tilde x) = \s (t,x)
\]
so that the estimate for $\tilde \s$, which we have established,
implies the desired estimate for $\nn{\delta (t)}_\infty$;
recall that we have been working in the transformed variables but
have dropped the tildas for the sake of convenience.
The proof of Theorem~\ref{stab} is now complete. \prfe 
In addition to the assertion in Theorem~\ref{stab} the following holds:
\begin{cor}
For every $\e_1 >0$ there exists $\e_0 >0$ such that every solution $g$
of $(\ref{pvlt}),\ (\ref{ppot}),\ (\ref{prhot})$ with $\gn \in \D$ and $\nn{\gn\,}_{1,\infty} < \e_0$
satisfies the estimate 
\[
\nn{\dx \s (t)}_\infty < a^{-2} (t) \e_1,\ t \geq 0 .
\]
\end{cor}
\prf 
By Lemma~\ref{sigdec} we obtain at the end of the proof of Theorem~\ref{stab} the estimate
\[
\nn{\dx \s (t)}_\infty \leq
C_1\, a^{-2} (t) \bigl( \nn{\gn\,}_{1,\infty} + \gamma \bigr)
< \epsilon_1 a^{-2} (t), 
\]
and the assertion follows. \prfe

\noindent
{\bf Final remarks:}
In the usual textbook explanations of the formation of galaxies it is 
argued that small perturbations of the homogeneous state lead to growth
in time of the density contrast $\delta (t,x)$ introduced in 
Theorem~\ref{stab}, cf.\ \cite{Pe1,Pe2,We}. However, we have proven that
$\nn{\delta (t)}_\infty$ can be made as small as we like, uniformly in
time, by making the initial perturbation small. This seeming 
contradiction can be resolved as follows: As already pointed out, our 
assumption $a(0) =1$ means that we perturb the homogeneous state
after is has already expanded out of the initial big bang singularity
for some time. If we would pose initial data for the perturbation
at times closer to the big bang then we would have to make
these perturbations smaller in order to retain the same bound on
$\nn{\delta (t)}_\infty$. One reason for this is that if $t_0 <0$
is defined by $a(t_0) =0$ then $a(t) \sim (t - t_0)^{2/3}$
for $t\geq t_0$ close to $t_0$, regardless whether the energy
$E_a$ of $a$ is positive, zero, or negative. For $E_a =0$ one has
$a(t) = C \,(t - t_0)^{2/3}$ for all $t \geq t_0$ in which case
the expansion would not be fast enough for our proof to go through.

On the other hand, precisely this case $E_a =0$ is considered
in the ``instability'' arguments in the textbooks which for the generic
case $E_a \neq 0$ is justified under the restriction to times close
to $t_0$, to which our analysis does not apply. 
Mathematically, this initial growth
of perturbations should thus not be referred to as an instability
since it is not an assertion on the behaviour of the perturbations
for all future time. 

After small perturbations have grown in the initial phase of the 
expansion the resulting solution, which on a larger scale can again 
be viewed as homogeneous, is now stable, provided we are in the 
case $E_a >0$,
and no further conglomeration beyond
the formation of galaxies and galaxy clusters occurs.

\end{document}